\documentclass[pra,showpacs,twocolumn,nofootinbib]{revtex4}
\usepackage{graphicx}
\usepackage{bm,amsmath}
\usepackage{dcolumn}
\usepackage{natbib}
\usepackage
{hyperref}

\newcommand{\fref}[1]{Fig.~\ref{#1}}

\newcommand{\sref}[1]{Sec.~\ref{#1}}
\newcommand{\Eref}[1]{Eq.~(\ref{#1})}
\newcommand{\tref}[1]{Table~\ref{#1}}
\newcommand{\rtw}{\longrightarrow}

\begin{document}

\title{Linear polyatomic molecules with $\Pi$ ground state: sensitivity to variation of
the fundamental constants}

\author{M. G. Kozlov}
\affiliation{
Petersburg Nuclear Physics Institute, Gatchina 188300, Russia}
\affiliation{
St.\ Petersburg Electrotechnical University ``LETI'', Russia}
\date{
\today}

\begin{abstract}
In polyatomic molecules with $\Pi$ electronic ground state the ro-vibrational
spectrum can be strongly modified by the Renner-Teller effect. The linear form
of C$_3$H molecule has particularly strong Renner-Teller interaction and a
very low lying vibronic $\Sigma^+_{1/2}$ level, which corresponds to the
excited bending vibrational mode. This leads to the increased sensitivities of
the microwave and submillimeter transition frequencies to the possible
variation of the fine structure constant $\alpha$ and electron to proton mass
ratio $\mu$.
\end{abstract}

\pacs{06.20.Jr, 06.30.Ft, 33.20.Bx}
\maketitle

\section{Introduction}\label{sec_intro}

At present it is generally recognized that microwave and submillimeter
molecular spectra  from the interstellar medium provide us with a very
sensitive tool to study possible variation of the fundamental constants
$\alpha=e^2/\hbar c$ and $\mu=m_e/m_p$ on a large space-time scale. It was
shown that certain types of transitions are particularly sensitive to such
variations. The mixed tunneling-rotational transitions in such molecules as
H$_3$O$^+$, H$_2$O$_2$, CH$_3$OH, and CH$_3$NH$_2$ can be very sensitive to
$\mu$-variation \cite{KPR11,Koz11,JXK11,JKX11,LKR11,IJKL12}. Recently the
submillimeter spectra of methanol have been used to place very stringent
limits on $\mu$-variation on the cosmological timescale \cite{LKR11,EVBL12}.
On the other hand, the $\Lambda$-doublet transitions in such diatomic
radicals, as OH and CH are very sensitive to variation of both constants
\cite{CK03,Dar03,Koz09}. The 18 cm OH line was observed at high redshifts,
which allowed to constrain time variation of both constants \cite{KLS12}. In
that work the 21 cm hyperfine hydrogen line was used as a reference. This
constrain can be further improved if more than one $\Lambda$-doublet
transition in OH, or CH is detected.

A rather general way to look for the enhanced sensitivity to variation of the
fundamental constants is to search for the accidental degeneracy of the levels
of different nature. This approach works for very different systems from
nuclei, to atoms, and molecules (see, for example, the review \cite{CFK09} and
\cite{BU09}).

In this paper we want to draw attention to the microwave and submillimeter
spectra of the linear polyatomic radicals with nonzero electronic angular
momentum. First, these molecules have $K$-doublets, which are analogous to the
$\Lambda$-doublets in diatomics. Second, the Renner-Teller interaction can
lead to the anomalously low lying vibronic levels and cause enhanced
sensitivities of the mixed ro-vibronic transitions. Finally, there are many
linear polyatomic molecules, which are detected in the interstellar medium. In
this context one of the most interesting species is the linear C$_3$H
molecule, where the excited vibronic $\Sigma_{1/2}^+$ level lies less than 30
cm$^{-1}$ above the ground level $\Pi_{1/2}$ and where several mixed
transitions were recently measured in the molecular beam experiment
\cite{CGDM09}. Because of that we focus on this molecule here leaving other
similar molecules for a separate discussion.

The interstellar carbon-chain radicals of hydrocarbon series C$_n$H ($n =
2-6$) exist in linear and cyclic isomeric forms. Both forms are observed in
the millimeter-wave range toward dark and translucent molecular clouds and
circumstellar envelopes of carbon-rich stars
\cite{IFH84,JAE84,TGH85,YSO87,YSSD90,MW90,NOJ93,GLC93,THT00,CGK00,FCGC01,KOK04,PC07}.
A typical abundance of the linear radical $l$-C$_3$H,~--- the simplest odd
carbon chain radical under consideration in the present study,~--- is $\sim
10^{-9}$ relative to hydrogen. The cyclic-to-linear abundance ratio
[$c-$C$_3$H]/[$l$-C$_3$H]~$\sim 5-10$ in dark clouds \cite{THT00,FCGC01},
but decreases to $\sim 3$ in translucent clouds \cite{THT00}, and becomes less
than unity around carbon stars \cite{PC07}.
The cyclic and linear isomers of C$_3$H have also been detected in a
star-forming region \cite{KJ10} and in two extragalactic sources: the
star-burst galaxy NGC~253 \cite{MMM05} and the spiral galaxy located in front
of the quasar PKS~1830--211 at the redshift $z=0.89$ \cite{MBG11}. Thus,
$l$-C$_3$H lines have been detected in regions with kinetic temperature
ranging from $T_{\rm kin} \sim 10$~K (dark clouds) to several hundred Kelvin
(circumstellar envelopes, star-forming regions). The observed line intensities
are less or about 0.1~K.

The lines observed from the interstellar medium (ISM) are Doppler broadened,
so the linewidth $\Gamma\approx\Gamma_D=\omega\tfrac{\Delta V}{c}$, where
$\Delta V$ is the velocity distribution for the ISM and $c$ is speed of light.
This means that $\tfrac{\Gamma}{\omega}\approx\tfrac{\Delta V}{c}$
characterizes ISM and to a first approximation is independent on the frequency
of the transition $\omega$. Because of that for the astrophysical search of
the possible variation of the fundamental constants it is crucial to find
transitions with high dimensionless sensitivity coefficients defined as:
 \begin{align}\label{Qfactors}
  \frac{\delta\omega}{\omega}
  &=Q_\alpha\frac{\delta\alpha}{\alpha}
  +Q_\mu\frac{\delta\mu}{\mu}\,.
 \end{align}
In the optical waveband these sensitivity coefficients are typically of the
order of $10^{-2}$. In the microwave waveband they are typically of the order
of unity, but can be much bigger. Below we will calculate $Q$ factors for some
microwave and submillimeter transitions of the $l$-C$_3$H molecule and show
that they can reach the order $10^3$. As usual such enhanced sensitivities
take place for the low frequency transitions between quasi degenerate levels
of different nature.

\section{Renner-Teller effect}\label{sec_basics}

In this section we briefly recall the theory of the Renner-Teller effect in
polyatomic linear molecules \cite{Ren34,Hou62}. The total molecular angular
momentum $\bm{J}$ consists of several contributions. We have the overall
rotation of the nuclei $\bm{R}$. Then we have the vibrational angular momentum
$\bm{G}$ associated with the twofold degenerate bending vibration mode(s) and
the electronic angular momentum $\bm{L}$. Momentum $\bm{R}$ is perpendicular
to the molecular axis $\zeta$; two others have $\zeta$ projections $l$ and
$\Lambda$. We define momentum $\bm{N}=\bm{R}+\bm{G}+\bm{L}$ and its projection
$\langle N_\zeta\rangle =K=l+\Lambda$. Finally, we add electronic spin:
$\bm{J}=\bm{N}+\bm{S}$, $\langle J_\zeta\rangle=\Omega$.

Suppose we have $\Pi$ electronic state $|\Lambda=\pm 1\rangle$ and $v=1$
vibrational state of a bending mode $|l=\pm 1\rangle$. All together there are
4 states $|\Lambda=\pm 1\rangle |l=\pm 1\rangle$. We can rewrite them as one
doublet $\Delta$ state $|K=\pm 2\rangle$ and states $\Sigma^+$ and $\Sigma^-$.
In adiabatic approximation all four states are degenerate. \citet{Ren34}
showed that the states with the same quantum number $K=l+\Lambda$ strongly
interact, so $\Sigma^+$ and $\Sigma^-$ states repel from each other, while
$\Delta$ doublet in the first approximation remains unperturbed. We are
particularly interested in the case when one of the $\Sigma$ levels is pushed
close to the ground state $v=0$. This is what takes place in $l$-C$_3$H
molecule \cite{YSSD90,KYSO96,CGDM09}.

Consider linear polyatomic molecule with unpaired electron in the $\pi_\xi$
state in the molecular frame $\xi,\eta,\zeta$. Obviously, the bending energy
is different for bendings in $\xi\zeta$ and in $\eta\zeta$ planes:
$V_\pm=\tfrac12 k_\pm\chi^2$ (here $\chi$ is the supplement to the bond
angle). That means that the electronic energy depends on the angle $\phi$
between the electron and nuclear planes:
 \begin{align}\label{Hprim}
 H'=V'\cos 2\phi\,,
 \end{align}
where $2V'=V_+-V_-=k'\chi^2$. There is no reason for $V'$ to be small, so
$k'\sim k_\pm\sim 1$ a.u.\ and to a first approximation $k'$ does not depend
on $\alpha$ and $\mu$.

As long as interaction \eqref{Hprim} depends on the relative angle between
electron and vibrational rotation it changes angular quantum numbers as
follows: $\Delta\Lambda=-\Delta l=\pm2$ and $\Delta K=0$. This is exactly what
is necessary to produce splitting between $\Sigma^+$ and $\Sigma^-$ states
with $v=1$ as discussed above.

Interaction \eqref{Hprim} also mixes different vibrational levels with $\Delta
v=\pm2,\pm4,\dots$. Thus, we have, for example, the nonzero matrix element
(ME) $\langle 0,0,1,1|H'|2,2,-1,1\rangle$ between states
$|v,l,\Lambda,K\rangle$. Such mixings reduce effective value of the quantum
number $\Lambda$ and, therefore, reduce the spin-orbital splitting between
$\Pi_{1/2}$ and $\Pi_{3/2}$ states \cite{PMTM03},
 \begin{align}\label{Aeff}
 H_\mathrm{so} &\equiv A_\mathrm{eff}\Lambda\Sigma\,,
 \quad A_\mathrm{eff}=A\Lambda_\mathrm{eff}/\Lambda\,.
 \end{align}

Let us define the model more accurately. Following \cite{PMTM03} we write
Hamiltonian as:
 \begin{align}\label{Ham1}
 H &= H_e + T_v + A L_\zeta S_\zeta\,.
 \end{align}
Here ``electronic'' part $H_e$ includes all degrees of freedom except for the
bending vibrational mode and spin. For $l$-C$_3$H there are two bending modes,
but for simplicity we include second bending mode in $H_e$ too. Electronic MEs
in the $|\Lambda\rangle$ basis have the form:
 \begin{subequations}\label{Ham2}
 \begin{align}
 \label{Ham2a}
 \langle \pm1|H_e|\pm1 \rangle &= \frac{V_+ + V_-}{2} = \frac{k}{2}\chi^2\,,
 \\ \label{Ham2b}
 \langle \pm1|H_e|\mp1 \rangle &= \frac{k'}{2}\chi^2 \exp{(\mp 2i\phi)} \,.
 \end{align}
 \end{subequations}
Here $\chi$ and $\phi$ are vibrational coordinates for the bending mode.
Kinetic energy in these coordinates has the form:
 \begin{align}
 \label{Kin}
 T_v &= -\frac{1}{2MR^2} \left(\frac{\partial^2}{\partial\chi^2}
 +\frac{1}{\chi}\frac{\partial}{\partial\chi}
 +\frac{1}{\chi^2}\frac{\partial^2}{\partial\phi^2}
 \right)\,.
 \end{align}

We can use the basis set of 2D harmonic functions in polar coordinates
$\rho=\chi R$ and $\phi$ for the mass $M$ and force constant $k$:
 \begin{align}
 \label{Oscil1}
 \psi_{v,l}(\rho,\phi) &= R_{v,l}(\rho)\frac{1}{\sqrt{2\pi}}\exp{(il\phi)}\,.
 \end{align}
It is important that radial functions are orthogonal only for the same $l$:
 \begin{align}
 \label{Oscil2}
 \langle R_{v',l}|R_{v,l}\rangle = \delta_{v',v}\,.
 \end{align}
This allows for the nonzero MEs between states with different quantum number
$l$.
Averaging operator \eqref{Ham1} over vibrational functions we get:
 \begin{multline}\label{Ham4}
 \langle v',l'|H_e+T_v|v,l\rangle
 = \bigl[\omega_v(v+1)+A\Lambda S_\zeta \bigr] \delta_{v',v} \delta_{l',l}
 \\
 +\frac12 \langle R_{v'l'}|k'\chi^2|R_{vl}\rangle
 \exp{(\mp 2i\phi)} \delta_{l',l\pm2}
  \,.
 \end{multline}
The exponent here ensures the selection rule $\Lambda'=\Lambda\mp2$ for the
quantum number $\Lambda$ when we calculate MEs for the rotating molecule.

We solved eigenvalue problem for Hamiltonian \eqref{Ham1} using the basis set
of the 2D-harmonic oscillator. Matrix elements were formed according to
\Eref{Ham4}. As discussed above we neglected one of the bending modes leaving
only the one that produces $K=0$ level close to the ground state doublet
$K=1,\,\Omega=1/2,3/2$. Our model Hamiltonian has only 3 parameters, namely
$\omega_v$, $A$, and the dimensionless Renner-Teller parameter $\cal E$:
$k'={\cal E}k$. In Ref.\ \cite{PMTM03} the following values were obtained:
 \begin{align}
 \label{param1}
 &\omega_v = 589\,\, \mathrm{cm}^{-1},
 \quad
 A = 29\,\, \mathrm{cm}^{-1},
 \quad
 {\cal E} = 0.883\,.
 \end{align}
We fixed the values for $\omega_v$ and $A$ and varied the Renner-Teller
parameter $\cal E$ to fit five lowest levels for the given bending mode:
$\Pi_{1/2}$, $\Pi_{3/2}$, $\Sigma_{1/2}$, $\Delta_{3/2}$, and $\Delta_{5/2}$.
The optimal value appeared to be ${\cal E}=0.788$. The difference with
\eqref{param1} is probably due to the neglect of the anharmonic corrections
and second bending mode.

Our results are presented in \tref{tab1}. The first two columns give nominal
vibrational quantum number $v$ and its actual average value. We see that
Renner-Teller term in \eqref{Ham4} strongly mixes vibrational states. This
mixing also affects $\langle\Lambda\rangle$ and decreases spin-orbital
splittings as explained by \Eref{Aeff}.

\begin{table}[htb]
  \caption{Low lying energy levels for the bending mode $\omega_v=589$ cm$^{-1}$
  and their sensitivities $q_\alpha$ and $q_\mu$ to the variation of $\alpha$
  and $\mu$ respectively. All values are in cm$^{-1}$.}
  \label{tab1}
 \begin{tabular}{clcddcddcd}
 \hline\hline
 \multicolumn{1}{c}{$v_\mathrm{nom}$}
 &\multicolumn{1}{c}{$\langle v\rangle$}
 &\multicolumn{1}{c}{$K$}
 &\multicolumn{1}{c}{$\Omega$}
 &\multicolumn{1}{c}{$\langle\Lambda\rangle$}
 &\multicolumn{1}{c}{$E$}
 &\multicolumn{1}{c}{$\Delta$}
 &\multicolumn{1}{c}{\cite{PMTM03}}
 &\multicolumn{1}{c}{$q_\mu$}
 &\multicolumn{1}{c}{$q_\alpha$}
 \\
 \hline
  0 & 1.22  & 1 & 0.5 & 0.50 & 367.9 &  0.0   &  0.0 & 187.8  &-14.6  \\
  0 & 1.35  & 1 & 1.5 & 0.46 & 381.9 & 13.9   & 14.0 & 187.8  & 13.3  \\
  1 & 2.32  & 0 & 0.5 &-0.01 & 394.2 & 26.3   & 27.0 & 197.3  & -0.4  \\
  1 & 3.57  & 2 & 1.5 & 0.21 & 597.7 &229.7   &226.0 & 300.3  & -6.1  \\
  1 & 3.65  & 2 & 2.5 & 0.19 & 603.5 &235.5   &232.0 & 300.3  &  5.5  \\
 \hline\hline
\end{tabular}
\end{table}

The last two columns in \tref{tab1} give sensitivity coefficients $q_\alpha$
and $q_\mu$ in cm$^{-1}$:
  $$ \delta E = q_\alpha \frac{\delta \alpha}{\alpha}
  + q_\mu \frac{\delta \mu}{\mu}\,.$$
To get them we assumed that parameters \eqref{param1} scale in a following
way: $\omega_v\sim \mu^{1/2}$, $A\sim\alpha^2$, and $\cal E$ does not depend
on $\alpha$ and $\mu$. The dimensionless sensitivity coefficients
\eqref{Qfactors} for the transitions $\omega_{i,k}=E_k-E_i$ can be found as:
 $$Q_{i,k}=(q_k-q_i)/\omega_{i,k}\,.$$
In \tref{tab2} these coefficients are calculated for the same set of
parameters as in \tref{tab1} and for the slightly different parameters which
better fit experimental frequencies from Ref.\ \cite{CGDM09}. We see that
$Q$-factors are practically the same for both sets.

\begin{table}[htb]
  \caption{$Q$-factors for the transitions between states from \tref{tab1}
  and for parameters $A_\mathrm{eff}$ and $\Delta E_{\Sigma\Pi}$ defined
  by \eqref{Aeff} and \eqref{DelE} respectively. Frequencies are in cm$^{-1}$.}
  \label{tab2}
 \begin{tabular}{cdcddddddd}
 \hline\hline
 &&&&\multicolumn{3}{c}{Fit to  Ref.\ \cite{PMTM03}}
 &\multicolumn{3}{c}{Fit to  Ref.\ \cite{CGDM09}}\\
 \multicolumn{1}{c}{$K$}
 &\multicolumn{1}{c}{$\Omega$}
 &\multicolumn{1}{c}{$K'$}
 &\multicolumn{1}{c}{$\Omega'$}
 &\multicolumn{1}{c}{$\omega$}
 &\multicolumn{1}{c}{$Q_\mu$}
 &\multicolumn{1}{c}{$Q_\alpha$}
 &\multicolumn{1}{c}{$\omega$}
 &\multicolumn{1}{c}{$Q_\mu$}
 &\multicolumn{1}{c}{$Q_\alpha$}
 \\
 \hline
  1 & 0.5  & 1 & 1.5 &  13.9 & 0.00&  2.00 &  14.4 & 0.00&  2.00  \\
  1 & 1.5  & 0 & 0.5 &  12.4 & 0.78& -1.11 &  13.3 & 0.77& -1.07  \\
  0 & 0.5  & 2 & 1.5 & 203.5 & 0.51& -0.03 & 204.4 & 0.51& -0.03  \\
  2 & 1.5  & 2 & 2.5 &   5.8 & 0.00&  2.00 &   6.0 & 0.00&  2.00  \\
 \multicolumn{4}{c}{$A_\mathrm{eff}$}
                     &  13.9 & 0.00&  2.00 &  14.4 & 0.00&  2.00  \\
 \multicolumn{4}{c}{$\Delta E_{\Sigma\Pi}$}
                     &  19.4 & 0.50&  0.00 &  20.5 & 0.50&  0.00  \\

 \hline\hline
\end{tabular}
\end{table}

For the two fine structure transitions, $\Pi_{1/2}\rtw \Pi_{3/2}$ and
$\Delta_{3/2}\rtw \Delta_{5/2}$, we get sensitivities $Q_\mu=0$ and
$Q_\alpha=2$. This may seem strange as the fine structure is significantly
reduced by the Renner-Teller mixing: the fine-structure parameter is 29
cm$^{-1}$ and the splitting between $\Pi_{1/2}$ and $\Pi_{3/2}$ is only 13.9
cm$^{-1}$. According to \eqref{Aeff} the mixing reduces the splitting.
However, this effect depends on the dimensionless Renner-Teller parameter
$\cal E$ and does not depend on $\mu$ and $\alpha$. Consequently, the
effective parameter $A_\mathrm{eff}$ depends on the fundamental constants in
the same way as initial parameter $A$.

For the high frequency transition $\Sigma_{1/2}\rtw \Delta_{3/2}$, where
spin-orbital energy can be neglected, we get $Q_\mu=0.5$ and $Q_\alpha=0$.
These results are expected, because our model has only two dimensional
parameters: vibrational frequency, which is proportional to $\mu^{1/2}$ and
the fine structure parameter $A$, which scales as $\alpha^2$. Even though our
vibrational spectrum is far from that of a simple harmonic oscillator, the
non-diagonal MEs \eqref{Ham4} of the Hamiltonian \eqref{Ham1} still scale as
$\mu^{1/2}$. Therefore, if we neglect spin-orbital splittings, we get
$Q_\mu=1/2$ for all transitions. The only transition in \tref{tab2} where
spin-orbital energy and vibrational energy are close to each other is the
$\Pi_{3/2}\rtw \Sigma_{1/2}$ transition. Resultant frequency is roughly half
of the vibrational energy difference between $\Pi$ and $\Sigma$ states. This
leads to $Q_\mu\approx 1$ and $Q_\alpha\approx -1$.

The following analysis in \sref{analysis_rot} will be based on the effective
Hamiltonian for the rotating molecule. The latter includes only two parameters
from this section: the effective fine-structure parameter $A_\mathrm{eff}$ and
the energy difference between $\Sigma$ and $\Pi$ states,
 \begin{equation}\label{DelE}
 \Delta E_{\Sigma\Pi}=E(\Sigma^+)-\frac{E(\Pi_{1/2})+E(\Pi_{3/2})}{2}\,.
 \end{equation}
Numerical values for these parameter will be obtained from the fit to
experimental transition frequencies. Here we only need to determine the
dependence of these parameters on fundamental constants. \tref{tab2} shows
that $A_\mathrm{eff}\sim \alpha^2$ and $\Delta E_{\Sigma\Pi}\sim \mu^{1/2}$.
Once again, this is because the Renner-Teller mixing depends on the
dimensionless parameter $\cal E$ and \textit{does not} depend on $\alpha$ and
$\mu$.

\section{Effective Hamiltonian for rotating molecule}
\label{analysis_rot}

In this section we mostly follow Ref.\ \cite{CGDM09}. However, we prefer to
use the basis set for the Hund's case ``a'' as we did before in
\cite{Koz09,BKB11}. We define the effective Hamiltonian for the subspace of
the three lowest vibronic states $\Pi_{1/2}$, $\Pi_{3/2}$, and
$\Sigma^+_{1/2}$. We neglect hyperfine interaction and some minor centrifugal
corrections included in \cite{CGDM09}.

The basis ro-vibronic states for the Hund's case ``a'' have the form:
 \begin{equation*}
 |v,l,\Lambda,(K),S,\Sigma,J,\Omega,M\rangle
 =|v,l\rangle|\Lambda\rangle|S,\Sigma\rangle|J,\Omega,M\rangle\,.
 \end{equation*}
Here the quantum number $K$ does not appear explicitly, being defined as
$K=l+\Lambda$. From these basic states we form parity states as described in
\cite{BC03}:
\begin{widetext}
 \begin{align}
 \label{basis2}
 |\Pi_{1/2}\rangle
 &=|0,0,1,(1),\tfrac12,-\tfrac12,J,\tfrac12,M,p\rangle
 =\frac{1}{\sqrt{2}}|0,0\rangle
 \Bigl(|1\rangle
 |\tfrac12,-\tfrac12\rangle
 |J,\tfrac12,M\rangle
 +\chi_p |-1\rangle
 |\tfrac12,\tfrac12\rangle
 |J,-\tfrac12,M\rangle\Bigr) \,,
 \\
 \label{basis3}
 |\Pi_{3/2}\rangle
 &=|0,0,1,(1),\tfrac12,\tfrac12,J,\tfrac32,M,p\rangle
 =\frac{1}{\sqrt{2}}|0,0\rangle
 \Bigl(|1\rangle
 |\tfrac12,\tfrac12\rangle
 |J,\tfrac32,M\rangle
 +\chi_p |-1\rangle
 |\tfrac12,-\tfrac12\rangle
 |J,-\tfrac32,M\rangle\Bigr) \,,
 \\
 \label{basis4}
 |\Sigma^+_{1/2}\rangle
 &=|1,1,1,(0),\tfrac12,\tfrac12,J,\tfrac12,M,p\rangle
 =\frac{1}{2}\Bigl(|1,1\rangle|-1\rangle
 +|1,-1\rangle|1\rangle\Bigr)
 \Bigl(|\tfrac12,\tfrac12\rangle
 |J,\tfrac12,M\rangle
 +\chi_p |\tfrac12,-\tfrac12\rangle
 |J,-\tfrac12,M\rangle\Bigl) \,,
 \end{align}
\end{widetext}
where the parity dependent phase is $\chi_p=(-1)^{J-S}p$.

We can write rotational energy by adding vibrational angular momentum $\bm{G}$
to the usual expression:
 \begin{subequations}
 \label{Hrot}
 \begin{align}
 \label{Hrota}
 H_\mathrm{rot}&=B(\bm{J}-\bm{G}-\bm{L}-\bm{S})^2 = B[J(J+1)-\Omega^2]
 \\
 &-2B\sum_{q=\pm 1}
 [J_qG_q+J_qL_q+J_qS_q
 \nonumber\\
 &\qquad\qquad +G_qL_{-q}+G_qS_{-q}+L_qS_{-q}]
 \label{Hrotb}
 \\
 \label{Hrotc}
 &-B\sum_{q=\pm 1}
 [G_qG_{-q}+L_qL_{-q}+S_qS_{-q}]\,.
 \end{align}
 \end{subequations}
Here we use the recipe from \cite{BC03} that in the molecular frame all scalar
products involving total angular momentum $\bm{J}$ are written as $J_qX_q$
rather than $(-1)^qJ_qX_{-q}$. The last line \eqref{Hrotc} can be skipped as
it gives a constant independent of $J$, $\Omega$, and $p$. The terms in
\eqref{Hrotb} linear in $L_q$ turn to zero in the subspace $\Lambda=\pm1$. We
are left with the following operator for the rotational energy:
 \begin{multline}
 \label{Hrot2}
 H_\mathrm{rot}= B[J(J+1)-\Omega^2]- D[J(J+1)-\Omega^2]^2
 \\
 -2B\sum_{q=\pm 1}
 [J_qG_q+J_qS_q+G_qS_{-q}]\,,
 \end{multline}
where we added standard centrifugal correction to the main diagonal term.

It is straightforward to calculate MEs of this operator on the states
\eqref{basis2} -- \eqref{basis4}. The term $J_qS_q$ does not change quantum
number $l$ and can not mix $\Sigma$ and $\Pi$ states. The nonzero matrix
elements are:
 \begin{align}
 \label{ME1}
 \langle\Pi_{3/2}|\!-\!2BJ_qS_q|\Pi_{1/2}\rangle
 &= -B\sqrt{(J-\tfrac12)(J+\tfrac32)}\,,
 \\
 \label{ME2}
 \langle\Sigma^+_{1/2}|\!-\!2BJ_qS_q|\Sigma^+_{1/2}\rangle
 &= -B\chi_p(J+\tfrac12)\,.
 \end{align}
The operator $J_qG_q$ changes quantum number $l$ by one and mixes $\Sigma$ and
$\Pi$ states:
 \begin{align}
 \label{ME3}
 \langle\Sigma^+_{1/2}|\!-\!2BJ_qG_q|\Pi_{1/2}\rangle
 &= -\beta\chi_p(J+\tfrac12)\,,
 \\
 \label{ME4}
 \langle\Sigma^+_{1/2}|\!-\!2BJ_qG_q|\Pi_{3/2}\rangle
 &= \beta\sqrt{(J-\tfrac12)(J+\tfrac32)}\,,
 \end{align}
where $\beta$ is defined as
 \begin{align}\label{beta}
 \beta&=B\langle l=1|G_1|l=0\rangle\,.
 \end{align}
This ME can not be calculated within this formalism and is included as an
independent parameter of the effective Hamiltonian (see also
\sref{sec_scaling}). Finally, the term $G_qS_{-q}$ mixes $\Sigma$ and $\Pi$
states, but can not change the quantum number $\Omega$:
 \begin{align}
 \label{ME5}
 \langle\Sigma^+_{1/2}|\!-\!2BG_qS_{-q}|\Pi_{1/2}\rangle
 &= -\beta\,.
 \end{align}

In addition to the rotational energy the effective Hamiltonian must include
spin-orbit interaction \eqref{Aeff}, the energy splitting between $\Sigma$ and
$\Pi$ states $\Delta E_{\Sigma\Pi}$ and spin-rotation interaction. Following
\cite{BC03} we write the latter as:
 \begin{multline}\label{gamma}
 \gamma\, (\bm{N}\bm{S})
 =\gamma\, (\bm{J}-\bm{S})\bm{S}
 \\
 =\gamma \left(\!\Omega\Sigma+\!\!\!\sum_{q=\pm1} J_qS_q -S(S+1)\!\right).
 \end{multline}
The nontrivial part of this interaction is now reduced to the MEs \eqref{ME1}
and \eqref{ME2}.

Equations \eqref{ME3} -- \eqref{ME5} show that the Coriolis terms involving
vibrational angular momentum $\bm{G}$ lead to the $K$-doubling via interaction
between $\Pi$ and $\Sigma$ states. In contrast to the terms involving
electronic angular momentum $\bm{L}$ here we do not need mixing with excited
electronic states. Still, because of the relative smallness of the parameter
$\beta$ in \eqref{beta}, these latter terms can not be neglected. They have
exactly the same form as for diatomic molecules and are defined in Ref.\
\cite{BC03}.

Transition amplitudes between spin-rotational states of the $l$-C$_3$H
molecule are expressed through MEs of the dipole moment operator $\bm{D}$ on
the basic states \eqref{basis2} -- \eqref{basis4}. Generally speaking there
are both diagonal and nondiagonal MEs in vibrational quantum numbers $v,l$.
Let us estimate them using atomic units ($\hbar=m_e=|e|=1$).

In the molecular frame the diagonal ME is reduced to the dipole moment of the
molecule along the molecular axis $\langle v,l|D_\zeta|v,l\rangle\approx D$.
If we assume that the charge of the hydrogen atom in the molecule is $q$, then
$D\sim 2qR_0\sim 4q$, where $R_0$ is the bond length. Comparing this estimate
with calculated value $D=1.40$ \cite{Woo95} we get $q=0.35$. Now we can
estimate the nondiagonal ME: $\langle 0,0|D_1|1,-1\rangle \sim q \bar{\xi}\sim
q/\sqrt{M\omega_v}\sim qM^{-1/4}\sim 0.1q\sim 0.04$, where $\bar{\xi}$ is the
amplitude of the vibration and $M\sim 10^4$ is the reduced mass for this
vibration mode. We conclude that nondiagonal MEs are much smaller than
diagonal, so we will neglect them.

In this approximation we get the following expressions for the reduced MEs on
the basis states \eqref{basis2} -- \eqref{basis4}:
 \begin{multline}
 \label{E1a}
 \langle X_{\Omega},J',p'||D|| Y_{\Omega},J,p\rangle
 = \delta_{X,Y} (-1)^{J'-\Omega}
 \\
 \times\sqrt{(2J'+1)(2J+1)}
 \left(\begin{array}{ccc} J'& 1 & J\\ -\Omega & 0 &\Omega
       \end{array}\right) \frac{1-p'p}{2} D\,,
 \end{multline}
where $X$ and $Y$ denote either $\Pi$, or $\Sigma$ state. Below we use these
expressions and theoretical value $D=1.40$ a.u.\ \cite{Woo95} to estimate
reduced MEs for the microwave transitions in $l$-C$_3$H. The Einstein
coefficients $A$ for these transitions can be found as \cite{Sob79}:
 \begin{equation}\label{E1-A}
 A_{i\rightarrow j} =\frac{4\omega^3_{ij}}{3\hbar c}
 \frac{|\langle i||D|| j\rangle|^2 a_0^2}{2J_i+1}\,,
 \end{equation}
where reduced ME is in a.u.\ and $a_0$ is the Bohr radius.

\section{Scaling of the parameters of the effective Hamiltonian with $\alpha$ and $\mu$}
\label{sec_scaling}

Effective Hamiltonian described in \sref{analysis_rot} is essentially
equivalent to the one used in Ref.\ \cite{CGDM09}. We included centrifugal
corrections to most of the terms using the same definitions as in
\cite{CGDM09}. For the hyperfine structure we used usual parameters $a$, $b$,
$c$, and $d$. Note that in \cite{CGDM09} the constant $b_F=b+c/3$ was used
instead of $b$.

In this section we discuss how the parameters of the effective Hamiltonian
depend on the constants $\alpha$ and $\mu$ (see \tref{tab3}). The scaling of
the two largest parameters, $\Delta E_{\Sigma\Pi}\sim \mu^{1/2}$ and
$A_\mathrm{eff}\sim\alpha^2$, has been already discussed in \sref{sec_basics}.
The rotational constants $B_\Sigma$ and $B_\Pi$ linearly depend on $\mu$. The
spin-rotational interaction \eqref{gamma} appears from the second order cross
term in Coriolis and spin-orbit interactions, therefore $\gamma\sim
\alpha^2\mu$. For $\Pi$ states there are two additional terms of the
spin-rotational interaction with parameters $p$ and $q$. The first of them has
the same scaling, as $\gamma$, i.e.\ $p\sim\alpha^2\mu$. The second term is
quadratic in Coriolis interaction, so $q\sim\mu^2$. These scalings are obvious
from the expressions on pp.\ 362 and 531 of \cite{BC03}.

Let us now discuss the parameter $\beta$ defined by \eqref{beta}. It is
proportional to the nondiagonal ME $\langle 1|G_1|0\rangle$. According to Eq.\
(13) in \cite{Hou62}, the perpendicular component $G_1$ simultaneously depends
on the vibrational coordinates of the bending ($v_b$) and stretching ($v_s$)
modes. In the harmonic approximation it has nonzero MEs only between different
stretching vibrational states, i.e.:
 $$\langle v_b=1,v_s=1| G_1 |v_b=0,v_s=0\rangle\neq 0\,.$$
The ME in \eqref{beta} is diagonal in stretching quantum number $v_s$. It is
nonzero due to the anharmonic corrections which mix vibrational modes. Such
corrections appear in the first order of the adiabatic perturbation theory and
are proportional to the adiabatic expansion parameter $\mu^{1/4}$. Thus we can
expect that $\beta \sim 0.1B$. This estimate agrees well with the numerical
value obtained in \sref{sec_results2}. We conclude that $\beta\sim
\mu^{1/4}B\sim \mu^{5/4}$.

\begin{table}[htb]
\caption{Parameters of the effective rotational Hamiltonian and their scaling
with $\alpha$ and $\mu$.}
 \label{tab3}
\begin{tabular}{lddll}
\hline\hline\\[-6pt]
 \multicolumn{1}{c}{Param.}
 &\multicolumn{1}{c}{This work}
 &\multicolumn{1}{c}{Ref.\ \cite{CGDM09}}
 &\multicolumn{1}{c}{Units}
 &\multicolumn{1}{l}{Scaling}\\
 \hline
 $\Delta E_{\Sigma\Pi}$ &  609.9811     &  609.9742     & GHz &$ \alpha^0\mu^{1/2}$ \\
 $B_\Sigma$             &  11.2124327   &  11.2126703   & GHz &$ \alpha^0\mu^1$ \\
 $D_\Sigma$             &   4.548       &   4.867       & kHz &$ \alpha^0\mu^2$ \\
 $\gamma_\Sigma$        &  -35.800      &  -35.525      & MHz &$ \alpha^2\mu^1$ \\
 $\gamma_{\Sigma,D}$    &   18.04       &    0.549      & kHz &$ \alpha^2\mu^2$ \\
 $b_\Sigma$             &   -6.3        &   -6.29       & MHz &$ \alpha^2\mu^1$ \\
 $c_\Sigma$             &   31.8        &   27.17       & MHz &$ \alpha^2\mu^1$ \\
 $A_\mathrm{eff}$       &  432.7762     &  432.7898     & GHz &$ \alpha^2\mu^0$ \\
 $B_\Pi$                &  11.1892055   &  11.1891033   & GHz &$ \alpha^0\mu^1$ \\
 $D_\Pi$                &  5.356        &  5.2340       & kHz &$ \alpha^0\mu^2$ \\
 $\gamma_\Pi$           &  -48.652      &  -48.075      & MHz &$ \alpha^2\mu^1$ \\
 $\gamma_{\Pi,D}$       &   21.670      &    0.000      & kHz &$ \alpha^2\mu^2$ \\
 $p$                    &  -6.9021      &  -7.0681      & MHz &$ \alpha^2\mu^1$ \\
 $p_D$                  & -1.595        &  0.504        & kHz &$ \alpha^2\mu^2$ \\
 $q$                    &  -12.8556     &  -12.9922     & MHz &$ \alpha^0\mu^2$ \\
 $q_D$                  &  -0.443       &  -0.1432      & kHz &$ \alpha^2\mu^3$ \\
 $\beta$                &  1.2586       &  1.2342       & GHz &$ \alpha^0\mu^{5/4}$ \\
 $\beta_D$              &  -28.3        &  -19.2        & kHz &$ \alpha^0\mu^{9/4}$ \\
 $a_\Pi$                &   12.43       &   12.32       & MHz &$ \alpha^2\mu^1$ \\
 $b_\Pi$                &  -22.57       &  -23.04       & MHz &$ \alpha^2\mu^1$ \\
 $c_\Pi$                &   27.56       &   28.07       & MHz &$ \alpha^2\mu^1$ \\
 $d_\Pi$                &   16.21       &   16.26       & MHz &$ \alpha^2\mu^1$ \\
\hline\hline
\end{tabular}
\end{table}

Our effective Hamiltonian includes centrifugal corrections ($D$, $\gamma_D$,
$\beta_D$, etc.) to the most important terms. We assume that such corrections
have the same $\alpha$ dependence as the respective main term and an extra
power in their $\mu$ dependence. The magnetic hyperfine constants scale as the
product of the nuclear and electronic magnetic moments, i.e. as $\alpha^2\mu$.

All scalings discussed above are approximate. There are relativistic
corrections to all parameters, which modify their $\alpha$-dependence. These
corrections are of the order of $(\alpha Z)^2 \sim 0.2\%$. The $\mu$
dependence of parameters is changed by non-adiabatic corrections. To
illustrate this point let us consider the rotational constants $B$. To a first
approximation the small difference between $B_\Sigma$ and $B_\Pi$ can be
related to the vibrational corrections to the adiabatic value of the
rotational constant $B_0$.

We can use the data from \tref{tab1} and \tref{tab3} to estimate vibrational
correction to the rotational constant:
 \begin{align}
 \label{rot1}
 B_v &= B_0 -\alpha(v+1)\,,\\
 \label{rot4}
 B_0 &= \frac{(v_\Sigma+1)B_\Pi-(v_\Pi+1)B_\Sigma}
   {v_\Sigma-v_\Pi}
   =11137.1\,\mathrm{MHz}\,,\\
 \label{rot5}
 \alpha &=\frac{B_\Pi-B_\Sigma}
   {v_\Sigma-v_\Pi}
   =-22.8\,\mathrm{MHz}\,.
 \end{align}
If we assume that $B_0$ scales as $\mu$ and $\alpha$ scales as $\mu^{3/2}$
\cite{LKR11}, we get following scalings of the rotational constants from
\tref{tab3}:
 \begin{align}
 \label{rot6}
 B_\Sigma \sim \mu^{1.010}\,,\qquad
 B_\Pi \sim \mu^{1.007}\,.
 \end{align}
Note that we neglected other vibrational degrees of freedom, so actual
corrections can be somewhat bigger. We conclude that we know the scalings of
the main parameters from \tref{tab3} roughly to a percent accuracy. Further
improvement of this accuracy requires extensive \textit{ab initio}
calculations.

\section{Numerical results for rotating molecule}\label{sec_results2}

\begin{figure}[tbh]
 \vspace*{-15mm}
 \includegraphics[scale=0.5]{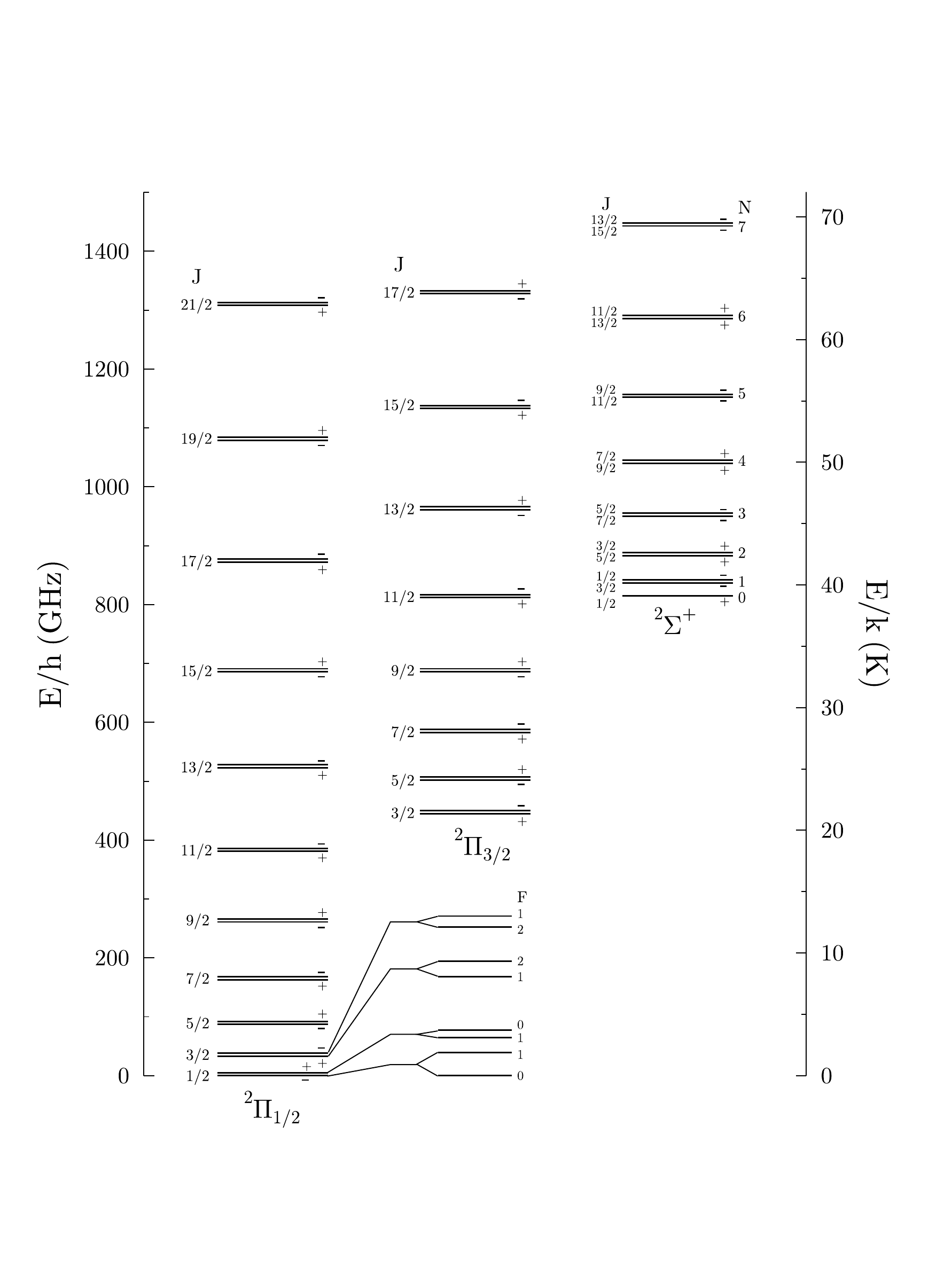}
 \vspace*{-20mm}
 \caption{Spin-rotational levels of the three lowest vibronic states of
the molecule $l$-C$_3$H. $K$-doubling is shown schematically. The levels are
labeled by the quantum numbers $J$ and $p$. The hyperfine structure for the
two lowest $K$-doublets is shown on the inset. Due to a strong Renner-Teller
effect the component $^2\Sigma^+$ of the excited bending state $\nu_4$(CCH
bending) is shifted towards lower energies, $\sim 29$ cm$^{-1}$ above the
zero-level of the ground state $^2\Pi_{1/2}$.}
 \label{fig_C3H}
\end{figure}

\begin{table}[hbt!]
\caption{Frequencies (MHz), $Q$-factors and reduced MEs (a.u.) of some
rotational transitions for $\Pi_{1/2}$, $\Pi_{3/2}$, and $\Sigma^+_{1/2}$
states.} \label{tabQ1-3}
\begin{tabular}{ccrrrd}
\hline\hline
 \multicolumn{6}{c}{$J \rightarrow J+1$ transitions for $\Pi_{1/2}$ state}
 \\
 \multicolumn{1}{c}{$J'\,p'$}
 &\multicolumn{1}{c}{$J\,p$}
 &\multicolumn{1}{c}{$\omega$}
 &\multicolumn{1}{c}{$Q_\alpha$}
 &\multicolumn{1}{c}{$Q_\mu$}
 &\multicolumn{1}{c}{$||D||^2$}
 \\
 \hline\\[-3mm]
  $\frac{ 3}{2}\,+$& $\frac{ 1}{2}\,-$&$  32627.84\ $& 0.06(0) &   0.97(1) &   1.86
  \\[1pt]
  $\frac{ 3}{2}\,-$& $\frac{ 1}{2}\,+$&$  32662.10\ $& 0.06(0) &   0.97(1) &   1.86
  \\[1pt]
  $\frac{ 5}{2}\,-$& $\frac{ 3}{2}\,+$&$  54405.75\ $& 0.06(0) &   0.97(1) &   3.33
  \\[1pt]
  $\frac{ 5}{2}\,+$& $\frac{ 3}{2}\,-$&$  54428.34\ $& 0.06(0) &   0.97(1) &   3.33
  \\[1pt]
  $\frac{ 7}{2}\,+$& $\frac{ 5}{2}\,-$&$  76199.10^\dagger
                                                    $& 0.06(0) &   0.97(1) &   4.72
  \\[1pt]
  $\frac{ 7}{2}\,-$& $\frac{ 5}{2}\,+$&$  76204.62^\dagger
                                                    $& 0.06(0) &   0.97(1) &   4.72
  \\[1pt]
  $\frac{35}{2}\,-$& $\frac{33}{2}\,+$&$ 383435.12^*$& 0.02(0) &   0.99(1) &  18.51
  \\[1pt]
  $\frac{35}{2}\,+$& $\frac{33}{2}\,-$&$ 383942.45^*$& 0.02(0) &   0.99(1) &  18.72
  \\[1pt]
  $\frac{47}{2}\,-$& $\frac{45}{2}\,+$&$ 516312.64^*$& 0.01(0) &   0.99(1) &  22.39
  \\[1pt]
  $\frac{49}{2}\,-$& $\frac{47}{2}\,+$&$ 539280.62^*$& 0.01(0) &   0.99(1) &  23.43
  \\[1pt]
 \hline
 \multicolumn{6}{c}{$J \rightarrow J+1$ transitions for $\Pi_{3/2}$ state}
 \\
 \multicolumn{1}{c}{$J'\,p'$}
 &\multicolumn{1}{c}{$J\,p$}
 &\multicolumn{1}{c}{$\omega$}
 &\multicolumn{1}{c}{$Q_\alpha$}
 &\multicolumn{1}{c}{$Q_\mu$}
 &\multicolumn{1}{c}{$||D||^2$}
 \\
 \hline\\[-3mm]
  $\frac{ 5}{2}\,-$& $\frac{ 3}{2}\,+$&$  57437.17\ $&$  -0.05(1) $& 1.03(1) &   2.22
  \\[1pt]
  $\frac{ 5}{2}\,+$& $\frac{ 3}{2}\,-$&$  57453.77\ $&$  -0.06(0) $& 1.03(1) &   2.22
  \\[1pt]
  $\frac{ 7}{2}\,+$& $\frac{ 5}{2}\,-$&$  80388.41\ $&$  -0.05(0) $& 1.03(1) &   3.93
  \\[1pt]
  $\frac{ 7}{2}\,-$& $\frac{ 5}{2}\,+$&$  80421.07\ $&$  -0.05(0) $& 1.03(1) &   3.93
  \\[1pt]
  $\frac{47}{2}\,-$& $\frac{45}{2}\,+$&$ 532658.83^*$&$  -0.06(0) $& 0.98(1) &  17.54
  \\[1pt]
  $\frac{49}{2}\,-$& $\frac{47}{2}\,+$&$ 556392.87\ $&$  -0.01(0) $& 1.00(1) &  23.16
  \\[1pt]
  $\frac{49}{2}\,+$& $\frac{47}{2}\,-$&$ 552385.85^*$&$  -0.12(0) $& 0.94(1) &  13.61
  \\[1pt]
  $\frac{51}{2}\,-$& $\frac{49}{2}\,+$&$ 599557.31\ $&$   0.09(1) $& 1.00(1) &  14.29
  \\[1pt]
  $\frac{51}{2}\,+$& $\frac{49}{2}\,-$&$ 578834.56\ $&$  -0.01(0) $& 1.00(1) &  23.77
  \\[1pt]
  $\frac{53}{2}\,-$& $\frac{51}{2}\,+$&$ 601263.86\ $&$  -0.01(0) $& 1.00(1) &  24.38
  \\[1pt]
 \hline
 \multicolumn{6}{c}{$N \rightarrow N+1$ transitions for $\Sigma^+_{1/2}$ state}
 \\
 \multicolumn{1}{c}{$N'\,J'\,p'$}
 &\multicolumn{1}{c}{$N\,J\,p$}
 &\multicolumn{1}{c}{$\omega$}
 &\multicolumn{1}{c}{$Q_\alpha$}
 &\multicolumn{1}{c}{$Q_\mu$}
 &\multicolumn{1}{c}{$||D||^2$}
 \\
 \hline\\[-3mm]
  $ 1\,\frac{ 1}{2}\,-$& $ 0\,\frac{ 1}{2}\,+$&$  22468.22\ $&$ 0.00(0) $& 1.00(1) &   0.93
  \\[1pt]
  $ 1\,\frac{ 3}{2}\,-$& $ 0\,\frac{ 1}{2}\,+$&$  22420.66\ $&$ 0.00(0) $& 1.00(1) &   1.87
  \\[1pt]
  $ 2\,\frac{ 3}{2}\,+$& $ 1\,\frac{ 1}{2}\,-$&$  44888.43\ $&$ 0.00(0) $& 1.00(1) &   1.87
  \\[1pt]
  $ 2\,\frac{ 5}{2}\,+$& $ 1\,\frac{ 3}{2}\,-$&$  44857.31\ $&$ 0.00(0) $& 1.00(1) &   3.36
  \\[1pt]
  $23\,\frac{47}{2}\,-$& $22\,\frac{45}{2}\,+$&$ 517808.87^*$&$ 0.05(0) $& 1.02(1) &  26.02
  \\[1pt]
  $24\,\frac{47}{2}\,+$& $23\,\frac{45}{2}\,-$&$ 538178.20\ $&$ 0.00(0) $& 1.00(1) &  32.57
  \\[1pt]
  $24\,\frac{49}{2}\,+$& $23\,\frac{47}{2}\,-$&$ 542975.34\ $&$ 0.11(0) $& 1.07(2) &  19.58
  \\[1pt]
  $25\,\frac{49}{2}\,-$& $24\,\frac{47}{2}\,+$&$ 560575.62\ $&$ 0.00(0) $& 1.00(1) &  33.93
  \\[1pt]
  $25\,\frac{51}{2}\,-$& $24\,\frac{49}{2}\,+$&$ 540680.86\ $&$-0.11(1) $& 1.01(2) &  18.11
  \\[1pt]
  $26\,\frac{53}{2}\,+$& $25\,\frac{51}{2}\,-$&$ 586696.44\ $&$-0.07(0) $& 0.94(1) &  23.27
  \\[1pt]
 \hline
 \end{tabular}
 \\
 $^\dagger$ Transitions detected at the redshift $z=0.89$ in Ref.\
 \cite{MBG11},\\
 $^*$ Transitions observed in Ref.\ \cite{CGDM09}.
 \end{table}

Our effective Hamiltonian has 22 parameters listed in \tref{tab3} including 6
parameters for the hyperfine structure. The 16 non-hyperfine parameters were
fitted using simplex method to the 44 experimentally observed transitions from
Ref.\ \cite{CGDM09} and to 12 experimental frequencies listed in the NIST
database \cite{NIST_Lovas}. We added 8 theoretically predicted transitions for
lower rotational quantum numbers from the same database to be sure we
adequately reproduce this part of the spectrum.

In our fit the rms deviation for 64 fitted transition is 0.23 MHz with maximum
deviation 0.52 MHz. This accuracy is lower than typical accuracy of the
similar fits in the literature, but is absolutely sufficient for our purposes.
Our main goal here is to calculate sensitivity coefficients for different
transitions to the variation of the fundamental constants. Though the
sufficiently complex effective Hamiltonians allow for very accurate
predictions of the transition frequencies, the accuracy they can provide for
the sensitivity coefficients is limited by the uncertainty in the dependence
of the used parameters on the fundamental constants (see \sref{sec_scaling}
and \cite{LKR11}).

To fit the hyperfine structure parameters we used 30 lines from the Ref.\
\cite{YSSD90} and 12 $K$-doublet transitions from \cite{JPL_Catalog}. The
hyperfine structure is mostly too small to change the values of the
sensitivity coefficients. This in not true only for several $K$-doublet
transitions with frequencies $\lesssim 100$~MHz, comparable to the hyperfine
splittings. We used the scalings from \tref{tab3} to calculate the shifts of
the spin-rotational levels due to the change of the constants $\alpha$ and
$\mu$ by $\pm0.1\%$. After that we found dimensionless sensitivities
$Q_\alpha$ and $Q_\mu$ for the transitions described by the effective
Hamiltonian.

\begin{table}[htb]
\caption{Frequencies (MHz), $Q$-factors and reduced MEs  (a.u.) for
$K$-doubling transitions in $\Pi_{1/2}$ state.} \label{tabQ4hfs}
\begin{tabular}{crddd}
\hline\hline
 \multicolumn{1}{c}{$J\,F'\!p'\!,Fp$}
&\multicolumn{1}{c}{$\omega$} &\multicolumn{1}{c}{$Q_\alpha$}
&\multicolumn{1}{c}{$Q_\mu$}
&\multicolumn{1}{c}{$||D||^2$}\\
\hline\\[-3mm]
$\frac{1}{2}\,{1+,\,\,0-} $&   52.37 &  0.66(2) & 1.7(2)   & 0.333 \\[1pt]
$\frac{1}{2}\,{0+,\,\,1-} $&   39.12 &  0.20(2) & 1.9(2)   & 0.333 \\[1pt]
$\frac{1}{2}\,{1+,\,\,1-} $&   34.93 & -0.02(2) & 2.0(2)   & 0.667 \\[3pt]
$\frac{3}{2}\,{1-,\,\,1+} $&   85.55 &  0.65(2) & 1.7(1)   & 0.166 \\[1pt]
$\frac{3}{2}\,{2-,\,\,1+} $&   78.60 &  0.55(2) & 1.7(1)   & 0.033 \\[1pt]
$\frac{3}{2}\,{1-,\,\,2+} $&   75.23 &  0.43(2) & 1.8(1)   & 0.033 \\[1pt]
$\frac{3}{2}\,{2-,\,\,2+} $&   68.29 &  0.30(2) & 1.8(1)   & 0.299 \\[3pt]
$\frac{5}{2}\,{2+,\,\,2-} $&  107.19 &  0.95(2) & 1.5(1)   & 0.132 \\[1pt]
$\frac{5}{2}\,{3+,\,\,2-} $&   98.97 &  0.89(2) & 1.5(1)   & 0.009 \\[1pt]
$\frac{5}{2}\,{2+,\,\,3-} $&   98.83 &  0.82(2) & 1.6(1)   & 0.009 \\[1pt]
$\frac{5}{2}\,{3+,\,\,3-} $&   90.61 &  0.75(2) & 1.6(1)   & 0.188 \\[3pt]
$\frac{7}{2}\,{3-,\,\,3+} $&  112.38 &  1.63(2) & 1.2(1)   & 0.105 \\[1pt]
$\frac{7}{2}\,{4-,\,\,4+} $&   96.07 &  1.56(2) & 1.2(1)   & 0.136 \\[3pt]
$\frac{9}{2}\,{4+,\,\,4-} $&   95.75 &  3.22(4) & 0.36(7)  & 0.086 \\[1pt]
$\frac{9}{2}\,{5+,\,\,5-} $&   79.63 &  3.45(4) & 0.23(7)  & 0.105 \\[3pt]
$\frac{11}{2}\,{5-,\,\,5+}$&   52.81 &  9.1 (6) &-2.6 (3)  & 0.072 \\[1pt]
$\frac{11}{2}\,{6-,\,\,6+}$&   36.85 & 12.1 (6) &-4.1 (3)  & 0.085 \\[3pt]
$\frac{13}{2}\,{6-,\,\,6+}$&   20.25 &-34.  (2) &19.  (2)  & 0.062 \\[1pt]
$\frac{13}{2}\,{7-,\,\,7+}$&   36.06 &-18.  (2) &11.  (2)  & 0.071 \\[3pt]
$\frac{15}{2}\,{7+,\,\,7-}$&  126.59 & -7.6 (2) & 5.8 (4)  & 0.054 \\[1pt]
$\frac{15}{2}\,{8+,\,\,8-}$&  142.24 & -6.5 (2) & 5.3 (4)  & 0.061 \\[1pt]
$\frac{17}{2}\,{8-,\,\,8+}$&  268.76 & -4.7 (1) & 4.4(3)   & 0.047 \\[1pt]
$\frac{17}{2}\,{9-,\,\,9+}$&  284.25 & -4.3 (1) & 4.2(3)   & 0.053 \\[1pt]
$\frac{19}{2}\,{9+,\,\,9-}$&  448.75 & -3.59(7) & 3.8(3)   & 0.042 \\[1pt]
$\frac{19}{2}\,{10+,\,\,10-}$&464.07 & -3.39(7) & 3.7(3)   & 0.046 \\[1pt]
$\frac{21}{2}\,{10-,\,\,10+}$&668.02 & -2.97(6) & 3.5 (3)  & 0.038 \\[1pt]
$\frac{21}{2}\,{11-,\,\,11+}$&683.18 & -2.85(6) & 3.4 (3)  & 0.041 \\[1pt]
%
\hline
\end{tabular}
\end{table}

There are three manifolds of levels, which belong to the vibronic states
$\Pi_{1/2}$, $\Pi_{3/2}$, and $\Sigma_{1/2}^+$ (see \fref{fig_C3H}). According
to \eqref{E1a} the strongest transitions take place between levels of the same
manifold. The higher frequency transitions correspond to the change of the
rotational quantum number $J\rightarrow J+1$ (see Table \ref{tabQ1-3}). Such
transitions usually have $Q_\alpha\approx 0$, $Q_\mu\approx 1$ \cite{dNWB12}.
We see that this is also true for $l$-C$_3$H.

For the $\Pi_{1/2}$ and $\Pi_{3/2}$ manifolds there is weak monotonic
dependence of the sensitivities on $J$. This dependence is caused by the
Coriolis interaction between these manifolds. For the upper part of the
$\Pi_{3/2}$ spectrum we see some irregularities in sensitivities. They are
caused by the resonant interactions with the nearby levels of the
$\Sigma_{1/2}$ manifold, where similar irregularities are observed for
$N\ge22$. All these irregularities are weak because interaction energy is much
smaller than respective transition frequencies.

\begin{table}[htb]
\caption{Frequencies (MHz), $Q$-factors and reduced MEs  (a.u.) for
$K$-doubling transitions in $\Pi_{3/2}$ state.} \label{tabQ5hfs}
\begin{tabular}{crddd}
\hline\hline
 \multicolumn{1}{c}{$J\,F'\!p'\!,Fp$}
 &\multicolumn{1}{c}{$\omega$}
 &\multicolumn{1}{c}{$Q_\alpha$} &\multicolumn{1}{c}{$Q_\mu$}
 &\multicolumn{1}{c}{$||D||^2$}\\
\hline\\[-3mm]
$\frac{3}{2}\,{1-,1+} $&     5.61 & -2.63(8) &  3.2 (2) & 1.493 \\[1pt]
$\frac{3}{2}\,{2-,1+} $&    18.50 &  0.49(8) &  1.7 (2) & 0.299 \\[1pt]
$\frac{3}{2}\,{1-,2+} $&    -7.30 &  5.28(8) & -0.6 (2) & 0.299 \\[1pt]
$\frac{3}{2}\,{2-,2+} $&     5.58 & -2.63(8) &  3.2 (2) & 2.688 \\[3pt]
$\frac{5}{2}\,{2+,2-} $&    22.24 & -2.60(8) &  3.2 (2) & 1.186 \\[1pt]
$\frac{5}{2}\,{3+,2-} $&    31.50 & -1.35(8) &  2.6 (2) & 0.085 \\[1pt]
$\frac{5}{2}\,{2+,3-} $&    12.88 & -5.67(8) &  4.6 (2) & 0.085 \\[1pt]
$\frac{5}{2}\,{3+,3-} $&    22.15 & -2.60(8) &  3.2 (2) & 1.694 \\[3pt]
$\frac{7}{2}\,{3-,3+} $&    54.92 & -2.57(8) &  3.2 (2) & 0.943 \\[1pt]
$\frac{7}{2}\,{4-,4+} $&    54.76 & -2.57(8) &  3.2 (2) & 1.223 \\[5pt]
$\frac{9}{2}\,{+-} $&  108.13 & -2.50(8) &  3.1 (2) & 1.230 \\[1pt]
$\frac{11}{2}\,{-+}$&  185.99 & -2.46(8) &  3.1 (2) & 1.007 \\[1pt]
$\frac{13}{2}\,{+-}$&  291.71 & -2.41(9) &  3.0 (2) & 0.847 \\[1pt]
$\frac{15}{2}\,{-+}$&  427.87 & -2.35(8) &  3.0 (2) & 0.727 \\[1pt]
$\frac{17}{2}\,{+-}$&  596.34 & -2.30(8) &  2.9 (2) & 0.633 \\[1pt]
$\frac{19}{2}\,{-+}$&  798.28 & -2.25(8) &  2.9 (2) & 0.558 \\[1pt]
$\frac{21}{2}\,{+-}$& 1034.16 & -2.21(9) &  2.9 (2) & 0.497 \\[1pt]
$\frac{23}{2}\,{-+}$& 1303.72 & -2.17(9) &  2.8 (2) & 0.446 \\[1pt]
$\frac{25}{2}\,{+-}$& 1605.97 & -2.15(9) &  2.8 (1) & 0.403 \\[1pt]
$\frac{27}{2}\,{-+}$& 1939.08 & -2.13(9) &  2.8 (1) & 0.366 \\[1pt]
$\frac{29}{2}\,{+-}$& 2300.16 & -2.13(9) &  2.7 (1) & 0.334 \\[1pt]
$\frac{31}{2}\,{-+}$& 2684.91 & -2.2 (1) &  2.7 (1) & 0.306 \\[1pt]
$\frac{33}{2}\,{+-}$& 3086.93 & -2.2 (1) &  2.7 (1) & 0.282 \\[1pt]
$\frac{35}{2}\,{-+}$& 3496.51 & -2.4 (1) &  2.6 (1) & 0.261 \\[1pt]
$\frac{37}{2}\,{+-}$& 3898.24 & -2.5 (1) &  2.60(9) & 0.242 \\[1pt]
$\frac{39}{2}\,{-+}$& 4266.17 & -2.9 (1) &  2.53(8) & 0.224 \\[1pt]
$\frac{41}{2}\,{+-}$& 4553.04 & -3.5 (1) &  2.42(5) & 0.208 \\[1pt]
$\frac{43}{2}\,{-+}$& 4663.43 & -4.6 (2) &  2.2 (1) & 0.192 \\[1pt]
$\frac{45}{2}\,{+-}$& 4377.16 & -7.5 (2) &  1.4 (3) & 0.174 \\[1pt]
$\frac{47}{2}\,{-+}$& 3097.96 &-19.0 (4) & -2.3 (9) & 0.149 \\[1pt]
$\frac{49}{2}\,{-+}$&  909.06 &132.  (2) & 53.(8)   & 0.103 \\[1pt]
$\frac{51}{2}\,{-+}$&19813.69 & -3.11(5) & -1.6 (4) & 0.116 \\[1pt]
$\frac{53}{2}\,{+-}$&16952.67 & -1.31(2) &  0.0 (4) & 0.136 \\[1pt]
$\frac{55}{2}\,{-+}$&16218.56 & -0.61(2) &  0.8 (4) & 0.138 \\[1pt]
\hline
\end{tabular}
\end{table}

For the $\Pi$ states there are also lower frequency transitions between the
levels of different parity with the same $J$ ($K$-doublets). For diatomic
radicals such transitions are known to be very sensitive to the variation of
both constants \cite{Dar03,CK03,Koz09}. Electron spin gradually decouples from
the molecular axis with growing rotational energy. As a result, the
$\Omega$-doubling for low $J$ values transforms to $\Lambda$-doubling for
higher $J$s. In our case the electronic quantum number $\Lambda$ is
substituted by the vibronic quantum number $K$, otherwise the effects are
rather similar (see Tables \ref{tabQ4hfs} and \ref{tabQ5hfs}). Decoupling of
the electron spin happens around $J=\tfrac{13}{2}$ and causes the anomaly in
sensitivities for the $\Pi_{1/2}$ doublets around $J=\tfrac{13}{2}$, where the
frequency drops below 50 MHz. For the $l$-C$_3$H molecule we can expect
additional anomalies in sensitivities due to the proximity and strong
interaction of $\Pi$ and $\Sigma$ states \cite{BKB11}. One such anomaly is
caused by the resonance between the levels $\Pi_{3/2}$ and $\Sigma^+_{1/2}$
with $J\approx\tfrac{49}{2}$. The transition frequency is higher here, about 1
GHz, but this is much smaller than for the neighboring rotational states.

The hyperfine structure is much larger for the $K$-doublets of the $\Pi_{1/2}$
state. For this reason we do not neglect the hyperfine structure in
\tref{tabQ4hfs}. For high $J$ values the transitions with $\Delta F\neq 0$ are
strongly suppressed, so we list only transitions with $\Delta F=0$. In
\tref{tabQ5hfs} the hyperfine splitting is neglected for all but the first few
transitions. For transitions with $J\ge \tfrac72$ the sensitivity coefficients
for the hyperfine components of the transition are practically the same.

\begin{table}[htb]
\caption{Frequencies (MHz), $Q$-factors and reduced MEs (a.u.) of some
transitions $\Pi_{1/2}\,J\,p \rightarrow \Pi_{3/2}\,J'\,p'$.} \label{tabQ6}
\begin{tabular}{ccrddd}
\hline\hline
 \multicolumn{1}{c}{$J'\,p'$}
 &\multicolumn{1}{c}{$J\,p$}
 &\multicolumn{1}{c}{$\omega$}
 &\multicolumn{1}{c}{$Q_\alpha$}
 &\multicolumn{1}{c}{$Q_\mu$}
 &\multicolumn{1}{c}{$||D||^2$}
 \\
 \hline\\[-3mm]
 $\frac{15}{2}\,-$& $\frac{17}{2}\,+$& 262072.96 &  3.00(2)  &  -0.50(0) &   0.41    \\[1pt]
 $\frac{17}{2}\,-$& $\frac{17}{2}\,+$& 456425.35 &  1.70(2)  &   0.15(0) &   1.20    \\[1pt]
 $\frac{19}{2}\,-$& $\frac{17}{2}\,+$& 674795.09 &  1.14(1)  &   0.43(1) &   0.87    \\[1pt]
 $\frac{15}{2}\,+$& $\frac{17}{2}\,-$& 261366.66 &  3.02(2)  &  -0.51(0) &   0.40    \\[1pt]
 $\frac{17}{2}\,+$& $\frac{17}{2}\,-$& 456743.26 &  1.70(2)  &   0.15(0) &   1.20    \\[1pt]
 $\frac{19}{2}\,+$& $\frac{17}{2}\,-$& 673718.38 &  1.14(1)  &   0.43(0) &   0.84    \\[1pt]
 $\frac{31}{2}\,-$& $\frac{33}{2}\,+$& 185259.51 &  3.45(2)  &  -0.71(1) &   2.31    \\[1pt]
 $\frac{33}{2}\,-$& $\frac{33}{2}\,+$& 558878.68 &  1.14(1)  &   0.43(1) &   5.46    \\[1pt]
 $\frac{35}{2}\,-$& $\frac{33}{2}\,+$& 961254.71 &  0.64(1)  &   0.68(1) &   3.30    \\[1pt]
 $\frac{31}{2}\,+$& $\frac{33}{2}\,-$& 179719.83 &  3.61(3)  &  -0.81(1) &   2.24    \\[1pt]
 $\frac{33}{2}\,+$& $\frac{33}{2}\,-$& 559110.84 &  1.13(1)  &   0.43(1) &   5.49    \\[1pt]
 $\frac{35}{2}\,+$& $\frac{33}{2}\,-$& 954903.42 &  0.66(1)  &   0.67(0) &   3.20    \\[1pt]
 $\frac{51}{2}\,-$& $\frac{53}{2}\,+$& 148111.71 &  2.91(2)  &  -1.01(6) &   3.92    \\[1pt]
 $\frac{53}{2}\,-$& $\frac{53}{2}\,+$& 729561.89 &  0.67(1)  &   0.67(0) &  12.80    \\[1pt]
 $\frac{55}{2}\,-$& $\frac{53}{2}\,+$&1369461.89 &  0.34(1)  &   0.82(1) &   6.63    \\[1pt]
 $\frac{51}{2}\,+$& $\frac{53}{2}\,-$& 118586.34 &  4.22(4)  &  -1.18(2) &   5.69    \\[1pt]
 $\frac{53}{2}\,+$& $\frac{53}{2}\,-$& 736802.86 &  0.64(1)  &   0.63(1) &  10.93    \\[1pt]
 $\frac{55}{2}\,+$& $\frac{53}{2}\,-$&1343531.64 &  0.36(1)  &   0.81(1) &   7.01    \\[1pt]
 \hline
 \end{tabular}
 \end{table}

\begin{table}[tbh]
\caption{Frequencies (MHz), $Q$-factors and reduced MEs (a.u.) of some
transitions $\Pi_{3/2}\,J\,p \rightarrow \Sigma^+_{1/2}\,N\,J'\,p'$. Negative
frequency means that final state lies lower.} \label{tabQ7}
\begin{tabular}{ccrddd}
\hline\hline
 \multicolumn{1}{c}{$N\,J'\,p'$}
 &\multicolumn{1}{c}{$J\,p$}
 &\multicolumn{1}{c}{$\omega$}
 &\multicolumn{1}{c}{$Q_\alpha$}
 &\multicolumn{1}{c}{$Q_\mu$}
 &\multicolumn{1}{c}{$||D||^2$}
 \\
 \hline\\[-3mm]
 $14\,\frac{29}{2}\,+$& $\frac{31}{2}\,-$&$-159987.95\   $&   2.00(2)  &   1.90(7) & 0.10 \\[1pt]
 $16\,\frac{31}{2}\,+$& $\frac{31}{2}\,-$&$ 535601.40^*  $&  -0.60(1)  &   0.73(1) & 0.14 \\[1pt]
 $16\,\frac{33}{2}\,+$& $\frac{31}{2}\,-$&$ 535512.07\   $&  -0.60(1)  &   0.73(1) & 0.11 \\[1pt]
 $16\,\frac{33}{2}\,+$& $\frac{33}{2}\,-$&$ 161892.90\   $&  -1.96(2)  &   0.12(7) & 0.21 \\[1pt]
 $15\,\frac{31}{2}\,-$& $\frac{33}{2}\,+$&$-200166.07\   $&   1.55(2)  &   1.74(6) & 0.14 \\[1pt]
 $17\,\frac{33}{2}\,-$& $\frac{33}{2}\,+$&$ 540214.36^*  $&  -0.58(1)  &   0.73(2) & 0.18 \\[1pt]
 $17\,\frac{35}{2}\,-$& $\frac{33}{2}\,+$&$ 540229.47\   $&  -0.57(1)  &   0.73(2) & 0.15 \\[1pt]
 $17\,\frac{35}{2}\,-$& $\frac{35}{2}\,+$&$ 144436.89\   $&  -2.13(2)  &  -0.01(8) & 0.28 \\[1pt]
 $16\,\frac{33}{2}\,+$& $\frac{35}{2}\,-$&$-240483.13\   $&   1.25(1)  &   1.63(6) & 0.19 \\[1pt]
 $18\,\frac{35}{2}\,+$& $\frac{35}{2}\,-$&$ 544660.22^*  $&  -0.55(1)  &   0.72(2) & 0.25 \\[1pt]
 $18\,\frac{37}{2}\,+$& $\frac{35}{2}\,-$&$ 544828.26\   $&  -0.55(1)  &   0.72(2) & 0.22 \\[-3pt]
 \quad\dots\\
 $23\,\frac{47}{2}\,-$& $\frac{49}{2}\,+$&$-514442.07\   $&     0.17(0)&    1.20(2)& 7.13 \\[1pt]
 $25\,\frac{49}{2}\,-$& $\frac{49}{2}\,+$&$ 579974.37\   $&    -0.24(1)&    0.79(2)&13.06 \\[1pt]
 $25\,\frac{51}{2}\,-$& $\frac{49}{2}\,+$&$ 569214.13^*  $&    -0.15(0)&    0.90(1)& 6.18 \\[1pt]
 $24\,\frac{47}{2}\,+$& $\frac{49}{2}\,-$&$  18489.70\   $&   -13.9 (2)&   -8.  (1)& 0.15 \\[1pt]
 $24\,\frac{49}{2}\,+$& $\frac{49}{2}\,-$&$  27624.21\   $&    -5.28(7)&   -3.0 (5)& 9.40 \\[1pt]
 $26\,\frac{51}{2}\,+$& $\frac{49}{2}\,-$&$1162035.24\   $&    -0.22(0)&    0.85(1)& 0.14 \\[1pt]
 $25\,\frac{49}{2}\,-$& $\frac{51}{2}\,+$&$    230.75\   $& -1099. (34)& -742. (90)& 0.17 \\[1pt]
 $25\,\frac{51}{2}\,-$& $\frac{51}{2}\,+$&$ -10529.49\   $&    18.8 (2)&   11.  (1)& 6.95 \\[1pt]
 $27\,\frac{53}{2}\,-$& $\frac{51}{2}\,+$&$1188561.69\   $&    -0.21(0)&    0.85(1)& 0.15 \\[1pt]
 $24\,\frac{49}{2}\,+$& $\frac{51}{2}\,-$&$-571024.04^*  $&     0.14(1)&    1.11(2)& 7.26 \\[1pt]
 $26\,\frac{51}{2}\,+$& $\frac{51}{2}\,-$&$ 563386.99\   $&    -0.34(1)&    0.79(3)&10.93 \\[1pt]
 $26\,\frac{53}{2}\,+$& $\frac{51}{2}\,-$&$ 556353.26\   $&    -0.32(1)&    0.84(2)& 5.39 \\[1pt]
 $26\,\frac{51}{2}\,+$& $\frac{53}{2}\,-$&$ -18063.18\   $&    13.6 (2)&   11.  (1)& 0.18 \\[1pt]
 $28\,\frac{53}{2}\,+$& $\frac{53}{2}\,-$&$ -25096.91\   $&     9.2 (1)&    6.7 (8)& 3.01 \\[1pt]
 $28\,\frac{55}{2}\,+$& $\frac{53}{2}\,-$&$1215046.58\   $&    -0.20(0)&    0.86(2)& 0.17 \\[1pt]
 \hline
 \end{tabular}
 \\
 $^*$ Transitions observed in Ref.\ \cite{CGDM09}.
 \end{table}

Because of the mixings \eqref{ME1} -- \eqref{ME5} of the basic states there
are also weaker transitions between $\Pi_{1/2}$, $\Pi_{3/2}$, and
$\Sigma^+_{1/2}$ manifolds. Examples of such transitions are listed in Tables
\ref{tabQ6}, \ref{tabQ7}. Sensitivities of these transitions depend on the
quantum numbers in a less regular manner, than sensitivities within each
manifold.

All transitions in \tref{tabQ6} have frequencies higher than 100 GHz. Because
of that the sensitivity coefficients are not very high, but they are dispersed
within intervals $0\lesssim Q_\alpha\lesssim 4$ and $-1\lesssim Q_\mu\lesssim
1$. Note that in order to study possible variations of fundamental constants
we need to compare several transitions with \textit{different} sensitivities.
Thus, such a spread in sensitivities can be very useful \cite{dNWB12}.

In \tref{tabQ7} there are several low frequency transitions with very high
sensitivities. Among them there are few with sufficiently high transition
amplitudes. In particular, there are three rather strong transitions at 27.6
GHz, 25.1 GHz, and 10.5 GHz with sensitivities $Q_\alpha$ from $-5$ to +19 and
$Q_\mu$ from $-3$ to +11. This is comparable to the sensitivities of the
$K$-doublet transitions from Tables \ref{tabQ4hfs} and \ref{tabQ5hfs}, but for
higher transition frequencies.

Transitions observed in \cite{CGDM09} are marked with asterisk in Tables
\ref{tabQ1-3} and \ref{tabQ7}. All of them have frequencies above 300 GHz and
sensitivities, which are not very far from the typical rotational
sensitivities: $Q_\alpha\approx 0$ and $Q_\mu\approx 1$. The maximal
difference in sensitivities $\Delta Q_\alpha \approx 0.7$ and $\Delta Q_\mu
\approx 0.4$ corresponds to the transitions at 535.6 and 571.0 GHz from
\tref{tabQ7}. The only two transitions, which were detected at high redshifts
in Ref.\ \cite{MBG11} are marked with the dagger in \tref{tabQ1-3}. These
transitions have much lower frequencies, but they are essentially rotational
transitions with ``normal'' sensitivities.

Let us discuss the accuracy of our calculations of the sensitivity
coefficients $Q_\alpha$ and $Q_\mu$. As we mentioned above, we know the
scalings of the parameters of the effective Hamiltonian with approximately 1\%
accuracy. So, we did several calculations of the sensitivity coefficients.
First, we changed all scalings by 1\%. Second, we used scaling of the
rotational constants from \Eref{rot6} keeping all other scalings unchanged.
The uncertainty in the scalings of the smaller parameters of the effective
Hamiltonian may be higher due to the nonadiabatic corrections. So we did two
additional calculations with the scaling of the parameter $\beta$ changed by
$\pm 1/4$, i.e. $\beta\sim \mu$ and $\beta\sim \mu^{3/2}$. Finally, in order
to check, how the fitting procedure may affect the results, we did several
calculations with slightly different sets of parameters. For example, we made
a 13 parameter fit with three centrifugal corrections set to zero:
$\gamma_{\Pi,D}=p_D=q_D=0$. In terms of the obtained frequencies, such fit is
only three times less accurate than our final 16 parametric one.

In Tables \ref{tabQ1-3} -- \ref{tabQ7} we give the average values of the $Q$
factors for all calculations, described above. The errors, given in the
brackets, correspond to the maximum deviations from these average values for
individual calculations. In most cases these errors are smaller than, or of
the order of 10\%, even for the large sensitivities. This accuracy is
sufficient for the analysis of the
experimental and observational data.\\

\section{Conclusion}

We have studied sensitivity coefficients to the variation of the fundamental
constants $\alpha$ and $\mu$ for the microwave and submillimeter spectra of
the linear polyatomic molecule with strong Renner-Teller interaction. As an
example we have chosen $l$-C$_3$H molecule, which is often observed in the
interstellar molecular clouds and recently has been detected at the redshift
$z=0.89$.

The Renner-Teller interaction depends on the dimensionless ratio ${\cal
E}=k'/k$ of the force constants in the two perpendicular planes, which include
molecular axis. Parameter $\cal E$ does not depend on the fundamental
constants and vibrational intervals scale in the same way as for harmonic
oscillator, i.e. $E_v\sim \mu^{1/2}$. However, the Renner-Teller interaction
modifies vibrational spectrum and can lead to the close lying vibrational
states. Such states then strongly interact with each other due to the Coriolis
interaction. As a result, the molecules with strong Renner-Teller interaction
can have low frequency mixed ro-vibronic transitions with strongly enhanced
sensitivity coefficients to the variation of $\alpha$ and $\mu$. For the
$l$-C$_3$H molecule we found several types of transitions with sensitivity
coefficients varying in a wide range. This opens new possibilities to study
variation of fundamental constants in astrophysics.

\acknowledgments I am grateful to Sergei Levshakov for bringing $l$-C$_3$H
molecule to my attention and for the constant interest to this work. I also
want to thank Ed Hinds, Vadim Ilyushin, and Wim Ubachs for valuable
discussions. This work is partly supported by the Russian Foundation for Basic
Research Grants No.\ 11-02-00943 and No.\ 11-02-12284-ofi-m and by The Royal
Society.


\end{document}